\documentstyle[preprint,aps,epsf,floats]{revtex}
\begin{document}
\tighten

\def\bfl{{\bbox \ell}}
\def\bull{\vrule height .9ex width .8ex depth -.1ex}
\def\Dslash{ {D\hskip-0.6em /} }
\def\MeV{{\rm MeV}}
\def\GeV{{\rm GeV}}
\def\Tr{{\rm Tr\,}}
\def\D{{\Delta}}
\def\a{{\alpha}}
\def\b{{\beta}}
\def\c{{\gamma}}
\def\d{{\delta}}
\def\m{{\mu}}
\def\M{{\cal M}}
\def\slash{{\!\not\!}}
\def\nrcpt{NR\raise.4ex\hbox{$\chi$}PT\ }
\def\ket#1{\vert#1\rangle}
\def\bra#1{\langle#1\vert}
\def\ltap{\ \raise.3ex\hbox{$<$\kern-.75em\lower1ex\hbox{$\sim$}}\ }
\def\gtap{\ \raise.3ex\hbox{$>$\kern-.75em\lower1ex\hbox{$\sim$}}\ }
\newcommand{\gsim}{\raisebox{-0.7ex}{$\stackrel{\textstyle >}{\sim}$ }}
\newcommand{\lsim}{\raisebox{-0.7ex}{$\stackrel{\textstyle <}{\sim}$ }}

\def\Journal#1#2#3#4{{#1} {\bf #2}, #3 (#4)}

\def\NCA{\em Nuovo Cimento}
\def\NIM{\em Nucl. Instrum. Methods}
\def\NIMA{{\em Nucl. Instrum. Methods} A}
\def\NPB{{\em Nucl. Phys.} B}
\def\NPA{{\em Nucl. Phys.} A}
\def\PLB{{\em Phys. Lett.}  B}
\def\PRL{\em Phys. Rev. Lett.}
\def\PRD{{\em Phys. Rev.} D}
\def\PRC{{\em Phys. Rev.} C}
\def\PRA{{\em Phys. Rev.} A}
\def\PR{{\em Phys. Rev.} }
\def\ZPC{{\em Z. Phys.} C}
\def\PREP{{\em Phys. Rep.}  }
\def\ANN{{\em Ann. Phys.} }
\def\SCI{{\em Science} }
\def\CJP{{\em Can. J. Phys.}}

\preprint{\vbox{
\hbox{INT00-83}
\hbox{NT@UW-00-03}
}}
\bigskip
\bigskip

\title{Meson Masses in High Density QCD}
\author{\bf Silas R. Beane$^a$,  
Paulo F. Bedaque$^b$, and
Martin J. Savage$^{a,c}$ }

\vspace{1cm}

\address{$^a$ Department of Physics 
\\ University of Washington, Seattle, WA 98195-1560}

\vspace{1cm}

\address{$^b$ Institute for Nuclear Theory
\\ University of Washington, Seattle, WA 98195-1560}

\vspace{1cm}

\address{$^c$ Jefferson Laboratory\\ 12000 Jefferson Avenue, Newport News, 
VA 23606}
\vspace{1cm}
\address{\tt sbeane, bedaque, savage@phys.washington.edu}

\maketitle

\begin{abstract}
The low-energy effective theories for the two- and three-flavor
color-superconductors arising in the  high density limit of QCD
are discussed. 
Using an effective field theory to describe  quarks near the 
fermi surface, we compute the masses of the pseudo-Goldstone
bosons that dominate the low-momentum dynamics of these systems.

\end{abstract}

\vfill\eject

\section{Introduction}

Recent developments have reinvigorated efforts to understand QCD at very high
baryon density~\cite{alford1}-\cite{Za}.  For special
combinations of quark colors and flavors it is likely that a
superconducting gap breaks color and flavor symmetries in interesting
ways. 
Although the symmetry breaking is nonperturbative, it occurs
when QCD is weakly coupled, and therefore perturbative QCD (pQCD) 
can be used to derive  properties of the superconducting
phase. 
One can conceive of scenarios in nature where it may be important to 
understand the behavior of QCD at high density:
for instance, in neutron stars and less likely, in
high energy heavy ion collisions.

Below the weak scale the standard model has the exact local gauge symmetries
$SU(3)_c\otimes U(1)_{\rm em}$ which describe the strong and
electromagnetic interactions.
In addition there is the exact global symmetry $U(1)_B$ corresponding to the 
conservation of baryon number 
and the approximate global symmetries
$SU(N_f)_L\otimes SU(N_f)_R\otimes U(1)_A$  for a theory with $N_f =2, 3$
light quarks.
These global symmetries are broken by the quark masses and 
in addition the $U(1)_A$ symmetry is also broken by  the strong anomaly.
At extremely high density, the contribution from the anomaly is suppressed
by powers of the chemical potential, $\mu$, and $U(1)_A$ is broken only by the 
light quark masses.
When a color-superconductor forms via the spontaneous breaking of color 
and flavor
symmetries, there will be pseudo-Goldstone bosons that contribute to 
or determine the very  low energy dynamics of such a system.
For sufficiently low energies it is clear that an effective field theory
description of these dynamics can be constructed, and will prove useful in 
computing
contributions to observables from the far-infrared region of the theory.
In recent work by Son and Stephanov\cite{SONSTE} and by 
Casalbuoni and Gatto\cite{CGa}
the masses and decay constants
of the pseudo-Goldstone bosons in the $N_f=3$ color-flavor scenario were 
computed
in the large-$\mu$ limit.  
It was found that the masses become independent of $\mu$, 
while the decay constants depend linearly on $\mu$.
Later work in \cite{dhong}\cite{RWZa} and \cite{ManTyt}
claim that the masses actually 
vanish
in the large-$\mu$ limit and are proportional to 
$\sim \D^2/\m^2 \log(\D/\m)$.
We agree with this later claim \cite{dhong,RWZa,ManTyt} 
and, through the use
of a hierarchy of effective theories, compute the  leading
contribution to the meson masses and decay constants.

\section{Three Flavors}

If we assume that the masses of the 
{\it up}, {\it down} and {\it strange} 
quarks are much smaller than the scale associated 
with the formation and dynamics of the color-superconducting state
(the gap $\D$), then it is
appropriate to consider a theory with three flavors of massless quarks, 
and include 
mass effects in perturbation theory.
In the limit of high densities the attraction leading to the gap is
given by
one-gluon exchange that is attractive in the color $\bar 3$ channel.
It has been argued that the most favorable state for the system
is one in which there is a formation of the ``color-flavor locked''
condensate
\begin{equation}
\langle \Psi^\a_{L a i} \Psi^\b_{L b j} \rangle = 
\langle \Psi^\a_{R a i} \Psi^\b_{R b j} \rangle = 
\D\epsilon^{\a \b  c} \epsilon_{abc} \epsilon_{ij} 
\end{equation}
\noindent
($\a,\b$,.. are color indices, $a,b,$..are flavor and $i,j$,... 
are spin indices)
resulting in the symmetry breaking pattern 
$SU(3)_c\otimes SU(3)_L\otimes SU(3)_R\otimes U(1)_A
\otimes U(1)_B\otimes U(1)_{\rm em}
\rightarrow SU(3)_{c+L+R}\otimes U(1)_{\rm \tilde{em}}$.
The value of the gap $\D$ was computed in 
\cite{SON1,HMSWa,SchaferWilczek,PisRa}
in the high density limit 
and does not have the usual BCS form but is instead given by
\begin{equation}
\Delta=c\  \frac{512}{g^5} \pi^4(\frac{2}{N_f})^{5/2}\mu 
  e^{-\frac{3\pi^2}
                 {\sqrt{2} g}
          }
\end{equation}
\noindent where $c$ is a constant of order unity not yet computed.
For lower densities, where pQCD does not apply, the same symmetry
breaking pattern was shown to occur by assuming that all quark interactions 
are effectively short ranged.
Throughout this work we assume that $\D$ is a constant, independent 
of energy and momentum.

As we are interested in modes close to the Fermi surface 
where $|p|\sim \mu \gg\D$ we can start by considering the dynamics of
quarks and gluons in the ungapped system,
where the dynamics are 
determined by the $3+1$ dimensional action, ${\cal S}_{\rm 3+1}$,
\begin{eqnarray} 
{\cal S}_{\rm 3+1}
& = & 
\int dt\ d^3x\ \left[\ 
-\frac{1}{4} G_{\mu\nu}^A G^{\mu\nu A}\  +\ 
\bar\psi_\a^a
(i\Dslash +\mu\gamma^0)^\a_\b \psi^{\b}_a
\ +\ \bar\psi_{L\a}^a\M_{a}^{b} \psi_{R b}^\a
\ +\ 
\bar\psi_{R\a}^a\M_{a}^{b\dagger} \psi_{L b}^\a
\ \right]
\ \ \ .
\label{qcd} 
\end{eqnarray}
\noindent
It is convenient to project onto the positive and negative energy
states, $\psi_+$ and $\psi_-$ respectively, and then eliminate
$\psi_-$ using the equations of motion\cite{HongEFT}, with 
\begin{eqnarray}
\psi & = & \psi_+\ +\ \psi_-
\ \ \ ,\ \ \ \psi_\pm\ =\ {\cal P}_\pm\psi
\ \ \ ,\ \ \ 
{\cal P}_\pm  =  
{1\over 2}\left( 1 \pm \gamma_0 {\bf\gamma}_k {\bf n}^k\right)
\ \ \ ,
\end{eqnarray}
where ${\bf n} = {\bf p}/|{\bf p}|$.
This procedure is similar to that used in the construction of
heavy quark effective theory (HQET)\cite{HQET}. 
Writing $\tilde{\cal S}_{\rm 3+1}$ in terms of the mode expansion
for $\psi_+$ and working in spherical coordinates,
the action describing the dynamics of the modes near the 
fermi surface, $\tilde {\cal S}_{\rm 3+1}$,  is
\begin{eqnarray}
\tilde{\cal S}_{\rm 3+1}
& = &  {\mu^2\over\pi}\ {\cal S}_{\rm 1+1}
\ \ \ ,
\end{eqnarray}
where $ {\cal S}_{\rm 1+1}$ is the action of a $1+1$ dimensional field theory.
The two-component quark field $\psi_+ = \chi^\alpha_a (E,k,{\bf n})$
depends on an energy $E\ll \mu$, a momentum $k\ll\mu$,
and a unit vector pointing toward the fermi surface ${\bf n}$.
As the anti-quarks 
have been integrated out of the theory at the scale $\mu$ corresponding to 
the top of the fermi surface, the effective field theory described by 
${\cal S}_{\rm 1+1}$
will be an expansion in terms of $E/\mu$ and $k/\mu$, as outlined in 
\cite{HongEFT}.

Perturbative computations around the superconducting state can be
performed  by adding and subtracting a quark gap term in the QCD
lagrangian. The condition that the subtracted gap term does not
contribute
at each order in perturbation theory is equivalent to the gap
equation.
The gaps for the positive and negative energy states will in general 
be different, and in terms of the 
$\psi_{\pm}$ fields we have an additional contribution to the lagrange
density of the form
\begin{eqnarray}
{\cal L}^\Delta & = & 
\frac{\D}{2}\epsilon_{\alpha\beta I } \epsilon^{abI} \psi^{T\a}_{a +} C
\psi^{\b}_{b +}
\ +\ 
\frac{\overline{\D}}{2}\epsilon_{\alpha\beta I } \epsilon^{abI} 
\psi^{T\a}_{a -} C
\psi^{\b}_{b -}
\ +\ {\rm h.c.}
\ \ \ .
\end{eqnarray}
The anti-gap, $\overline{\D}$, has not been computed at this point
in time,
but a recent discussion can be found in \cite{PisRa}.

A further simplification can be made by writing the quark fields, $\chi$, 
in terms of the mass eigenstates of the condensate (neglecting the 
quark masses)
\begin{eqnarray}
\chi^{\a}_a & = & {1\over\sqrt{2}} \sum_{A=1}^9 \chi^A
 \left(\lambda^A\right)^{\a}_a
\ \ \ ,
\end{eqnarray}
where $\lambda^A$ with $A=1,..,8$ are the Gell-Mann matrices and 
$\lambda^9=\sqrt{2/3}\ I_3$.
After eliminating $\psi_-$ by the equations of motion,
the part of the leading order action that
does not depend upon the quark masses is 
\begin{eqnarray}
{\cal S}_{\rm 1+1}^{(0)} & = & \sum_{A=1}^9
\int {\rm d {\bf n}\over 4\pi}
{\rm dE \ dk\over (2\pi)^2}
\left[\ 
\chi^{A\dagger}_{\bf n}
\left[ E - k
\right]\chi^{A}_{\bf n}
\ -\ {\Delta^A\over 2}\left( \chi^{A T}_{\bf n} C \chi^{A}_{-\bf n}
\ +\ {\rm h.c.}\right)
\ +\ \cdots
\right] 
\ \ \ ,
\label{eq:Lzero}
\end{eqnarray}
where $\Delta^A=\Delta$ for $A=1,..8$, while $\Delta^9=-2\Delta$.
Interactions between the $\chi^A$ and the gauge fields have not been shown.
The ellipses denote operators that are suppressed by powers of $\mu$.
As we are assuming that the quark masses are small compared to the 
scales associated with formation of the superconducting state,
we can treat the quark masses in perturbation theory.
The leading order contributions from the quark masses are described by the 
action
\begin{eqnarray}
{\cal S}_{\rm 1+1}^{(m^2)} & = &  -\sum_{A,B=1}^9
\int {\rm d {\bf n}\over 4\pi}
{\rm dE \ dk\over (2\pi)^2}
\left[\ 
{1\over 4\mu} 
\chi^{A\dagger}_{\bf n}\chi^{B}_{\bf n}
{\rm Tr} \left[\lambda^A {\cal M}^\dagger {\cal M}\lambda^B\right]
\right.\nonumber\\
& & \left.
\qquad
+ \frac{\overline{\D}}{16\mu^2}  
\left[ 
\left(\chi^{A\ T}_{R\bf n} C \chi^B_{R -\bf n} 
+ \chi^{A\dagger}_{L\bf n} C \chi^{B\ *}_{L-\bf n}\right)
{\cal Y}^{AB}
\ +\ {\rm h.c.}\right]
\ \right] 
\ \ \ ,
\label{eq:Lmtwo}
\end{eqnarray}
where
\begin{eqnarray}
{\cal Y}^{AB} & = & 
{\rm Tr}\left[ \lambda^A {\cal M}\lambda^B {\cal M}\right] 
\ -\   
{\rm Tr} \left[\lambda^A\M\right] 
\ \  {\rm Tr}\left[ \lambda^B {\cal M}\right]
\ \ \ ,
\end{eqnarray}
and where the $L,R$ subscripts on $\chi^A$ denotes the 
helicity/chirality state.
The appearance of explicit factors of $1/\mu$ associated with the mass terms
is no surprise, in fact, the first term in Eq.~(\ref{eq:Lmtwo})
follows naturally from the expansion of 
$E=\sqrt{p^2+m^2}=p + m^2/(2p)+..$ for $p\sim\mu\gg m$.
Contributions from more insertions of the light quark mass matrix or from
higher derivative operators are suppressed by powers of $\mu$.
The actions in Eq.~(\ref{eq:Lzero}) and  Eq.~(\ref{eq:Lmtwo})
describe the dynamics of modes near the fermi surface.
Contributions to observables arising from modes far from the fermi surface
are suppressed by powers of $\mu$, and enter through the higher dimension
operators that we have not shown.
An important point is that 
this lagrange density involves only analytic functions of 
the light  quark masses and the gap $\Delta$.   
For very low-energy dynamics of the system $|{\bf p}|\ll\Delta$
it is appropriate to construct 
an effective field theory for the pseudo-Goldstone modes alone.
This lagrange density will involve analytic functions of the 
light quark masses, but it  will have 
non-analytic dependence on $\Delta$.
This nonanalytic dependence can be computed from the effective theory 
describing the momentum region $\Delta \ll |{\bf p}| \ll \mu$, 
as described above.
This is in direct analogy with the nonanalytic contributions to observables
in the light meson sector, such as terms of the form 
$\log\left(m_\pi^2/\Lambda_\chi^2\right)$ or $\sqrt{m_q}$.

For momenta much below the gap,
$p\ll \D$,  the relevant degrees of freedom are
the nine pseudo-Goldstone bosons 
resulting from  symmetry breaking due to the condensate.
They are described by the fields 
\begin{eqnarray}
\Sigma & = & e^{i 2 M/ f_8}
\ \ ,\ \ 
V\ =\ e^{i 2 \eta_1/f_1}
\ \ ,
\end{eqnarray}
with
\begin{eqnarray}
M & = & 
\left(\matrix{ \pi_3/\sqrt{2} + \eta_8/\sqrt{6} & \pi^+ & K^+\cr
\pi^- & -\pi_3/\sqrt{2} + \eta_8/\sqrt{6} & K^0\cr
K^- & \overline{K}^0 & -2\eta_8/\sqrt{6} 
}\right)
\ \ \ .
\end{eqnarray}
These fields and the quark mass matrix transform under 
$SU(3)_L\times SU(3)_R\times U(1)_B\times
U(1)_A$ as
\begin{eqnarray}
\Sigma&\rightarrow & L\Sigma R^\dagger
\ \ ,\ \ V\rightarrow e^{-i4\beta}V
\ \ ,\ \ \M\rightarrow e^{+i2\beta}L\M R^\dagger
\ \ ,
\end{eqnarray}
\noindent
where $\beta$ is the $U(1)_A$ phase.
The interactions of these mesonic excitations will have an expansion
around zero momentum described by
\begin{eqnarray}
{\cal L} & = & 
{f_8^2\over 8}
{\rm Tr}\left[ \partial_0 \Sigma \partial_0 \Sigma^\dagger \right] 
\ -\ 
|{\bf v_8}|^2 
{f_8^2\over 8}
{\rm Tr}\left[ \nabla_i \Sigma \nabla_i \Sigma^\dagger \right] 
\ + \ {f_1^2\over 8}\ \partial_0 V \partial_0 V^* 
\ -\ 
|{\bf v_1}|^2
{f_1^2\over 8}
\ \nabla_i V \nabla_i V^*
\nonumber\\
&+&{A_1}\;\left( {\rm Tr}\; [\; \M^\dagger\;  \Sigma \;]
  \; {\rm Tr}\; [\; \M^\dagger\; \Sigma \;]\;  V^*
 + {\rm h.c.}\right)
\ +\ A_2\;
\left( {\rm Tr}\;[\; \M^\dagger\; \Sigma\;  \M^\dagger\; \Sigma \;]\; V^*
 + {\rm h.c.}\right)
\nonumber\\
& + & 
A_3\; {\rm Tr}\;[\; \M^\dagger\; \Sigma\; ]  
\; {\rm Tr}[\; \M\; \Sigma^\dagger\; ]
\ +\ A_4\   {\rm Tr}\;[\M \M^\dagger\ ]\ 
\ +\ \cdots
\ \ \ ,
\label{EFT}
\end{eqnarray}
\noindent
where the ellipses denote operators 
suppressed by powers of $p/\D$ and $\M/\D$
(for a recent discussion of this see \cite{Za}).
The velocities of the octet and singlet 
bosons are ${\bf v}_8$ and ${\bf v}_1$ respectively.
The coefficients appearing in Eq.(\ref{EFT}) are determined by 
matching S-matrix elements in the theory above $\Delta$ 
with those in the theory below $\Delta$, determined from 
Eqs.(\ref{eq:Lzero}), (\ref{eq:Lmtwo}) and Eq. (\ref{EFT}) respectively.
The form of the mass terms is exactly the same as the second order terms
appearing in the Lagrange density written down by 
Gasser and Leutwyler\cite{GLclassic}.

\subsection{Decay Constants}

The decay constants $f_8$ and $f_1$ 
can be found 
by computing the Debye and Meissner masses of fictitious 
gauge bosons coupled to 
currents in both 
pQCD \cite{SONSTE}\ and  the low-energy effective field theory, 
Eq.~(\ref{EFT}).
In the low-energy regime the two calculations must 
produce identical results,  
thereby determining 
$f_8$ and $f_1$
order by order in the chiral, $1/\mu$  and $\alpha_s$ expansions.

Gauging the $U(1)_A$ axial current leads to  
$D_\mu=\partial_\mu + i e W_\mu Q_1$, 
where $Q_1$ is the axial charge operator, 
$Q_1 \Psi_L = +1 \Psi_L$, $Q_1 \Psi_R = -1 \Psi_R$, $Q_1 V = -4 V$
and $Q_1\tilde\Sigma=0$.  It is then straightforward to show that the
mass of the $W_\mu$ fields at leading order 
in the low-energy effective field theory are 
\begin{eqnarray}
\Delta {\cal L}  & = & 2 \ e^2\  f_1^2\  
\left[\  W_0^2 - |{\bf v}_1|^2 W_i^2 \ \right]
\label{W0eft}
\ \ \ .
\end{eqnarray}
%
\begin{figure}[t]
\centerline{{\epsfxsize=4.4in \epsfbox{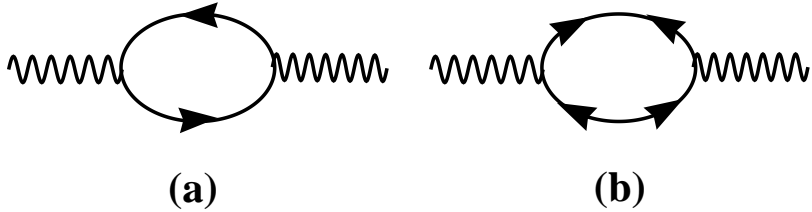}} }
\noindent
\caption{\it
One loop diagrams that give the leading contribution
to the meson decay constants.
}
\label{fig:decay}
\vskip .2in
\end{figure}
Computing these masses in pQCD through Eq.~(\ref{eq:Lzero})
one finds that the Debye mass
is given by the two graphs shown in Fig.~(\ref{fig:decay})
with the result
\begin{eqnarray}
\Delta {\cal L}_{\rm QCD} & = & e^2 {9\mu^2\over 2\pi^2}
\ W_0^2  
\ \ \ ,
\end{eqnarray}
and is
essentially identical 
to the standard many-body calculations  described in \cite{FWa}.
\noindent
In order to reproduce this with the effective theory
\begin{equation}
f_1=\frac{9}{4}\frac{\mu^2}{\pi^2}
\ \ \ ,
\end{equation}
which agrees with \cite{SONSTE}.
To compute the speed of the
Goldstone mode, we need to compute the Meissner mass in both
theories. 
In addition to the diagrams of Fig.~(\ref{fig:decay}) (with different 
couplings) 
there is a contribution from a counterterm.
The sum of diagrams yields a vanishing Meissner mass in the 
normal phase and a non-zero mass in the superconducting phase,
see \cite{FWa}.
The speed is found to be 
$|{\bf v}_1|^2 = {1\over 3}$,
in agreement with \cite{SONSTE}.
As discussed in \cite{SONSTE}
the analogous calculation for the  baryon number Goldstone boson 
is identical at leading order, 
$|{\bf v}_{H}|^2 = |{\bf v}_1|^2 = {1\over 3}$.

To determine the decay constant of the octet pseudo-Goldstone 
bosons, $f_8$, the Debye and Meissner masses
of fictitious gauge bosons associated with  octet currents
are computed.
We find  
\begin{eqnarray}
\label{piond}
f^2_8 & = &  {\mu^2\over\pi^2}\ 
{21-8\log [2]\over 9}
\ \ \ ,
\end{eqnarray}
which differs by a factor of 2 from Ref.~\cite{SONSTE} after differences
due to the definition of $f_8$ are taken into account.
The pion velocity is found to be 
$|{\bf v}_8|^2  =  {1\over 3}$
which agrees with Ref.~\cite{SONSTE}.  
Therefore, all Goldstone modes have the
same speed at leading order in the expansion.

As the decay constants scale like $f\sim\mu$, it is apparent that
the dynamics of the pseudo-Goldstone bosons do not receive
significant contributions from loops in the 
low-energy effective theory of Eq.~(\ref{EFT}).
The naive size of the counterterms is set by $1/\Delta$,
while the contribution from loops is set by $1/\mu$.
Therefore, once the coefficients in the low energy effective theory
have been determined, only tree-level diagrams need to be considered.

\subsection{Meson Masses}

The coefficients $A_i$
in Eq. (\ref{EFT}) can be determined by
matching the change in the ground state energy due to the 
quark masses in the high (Eqs. (\ref{eq:Lzero}) and (\ref{eq:Lmtwo}))
and low energy theories  (Eq. (\ref{EFT})).
In the low energy theory, the change in energy density can be easily 
determined from Eq. (\ref{EFT}) by setting $\Sigma = V I_3= I_3$,
to yield
\begin{eqnarray}
\delta{\cal E} & = & 
A_1 \left( \left({\rm Tr}\left[{\cal M}\right]\right)^2\ +\ {\rm h.c.}\right)
\ +\ 
A_2 \left( {\rm Tr}\left[{\cal M}^2\right] +\ {\rm h.c.}\right)
\nonumber\\
&  & +\  
A_3 {\rm Tr}\left[{\cal M}\right]{\rm Tr}\left[{\cal M}^\dagger\right]
\ +\ 
A_4 {\rm Tr}\left[{\cal M}{\cal M}^\dagger\right]
\ \ \ .
\end{eqnarray}
The operator with coefficient $A_4$ does not contribute to the dynamics
of the pseudo-Goldstone modes, and we do not calculate it.

Computation of the energy density in the $3+1$ dimensional high energy 
theory can easily be done, by noting that
the action of the $3+1$ dimensional theory is a factor of $\mu^2/\pi$
times the action of the $1+1$ dimensional theory.
Thus the energy density in the $3+1$ dimensional theory is $\mu^2/\pi$
times the energy density computed in the $1+1$ dimensional theory.
We use dimensional regularization and minimal subtraction
to define divergent integrals that 
occur at loop level in the $1+1$  dimensional theory.
\begin{figure}[t]
\centerline{{\epsfxsize=4.0in \epsfbox{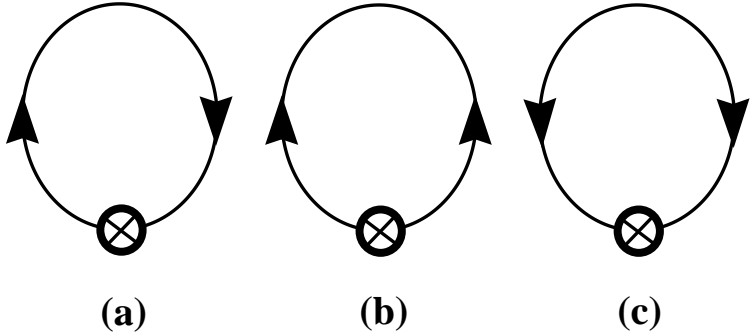}} }
\noindent
\caption{\it
One loop diagrams that give the leading contribution
to the meson masses.
Graph (a) contributes only to $A_4$ which does not impact the 
dynamics of the pseudo-Goldstone modes, while
graphs (b) and (c) contribute to both $A_1$ and $A_2$.
The crossed circle denotes an insertion of the quark
mass operators in the effective field theory defined in
Eq.~(\protect\ref{eq:Lmtwo}).
}
\label{fig:mass}
\vskip .2in
\end{figure}
The shift in the vacuum energy 
due to the light quark masses 
results from the  tadpole diagrams shown in
Fig.~(\ref{fig:mass}),
where the vertex arises from  Eq. (\ref{eq:Lmtwo}). 
At leading order we find 
 \begin{eqnarray}
 \delta {\cal E}^{\rm loop} & & = \frac{\overline{\D}\D}{4 \pi^2} 
 \log(\frac{\D}{\Lambda})
\  \left(\  \left({\rm Tr}\left[{\cal M}\right]\right)^2
\ -\ {\rm Tr}\left[{\cal M}^2\right]\ \right)
\ +\ {\rm h.c.}
\ \ \ ,
\label{E}
\end{eqnarray}
\noindent
where $\Lambda$ is the renormalization scale
and we have only shown the term nonanalytic in $\D/\Lambda$.
The form of our expression agrees with the results of
\cite{dhong,RWZa,ManTyt}.
The explicit dependence on $\Lambda$ shown in Eq.~(\ref{E})
is absorbed by an equal but opposite $\Lambda$ dependent 
counterterm that, for $\Lambda\sim \mu$
generates a shift in energy $\sim \overline{\D}\D/4 \pi^2$. 
This counterterm contribution
is suppressed compared to the contribution in Eq.(\ref{E}) 
by the large $\log\left( \Delta/\mu\right)$  factor and
will be neglected.
Matching this result with the corresponding shift in energy computed
in the effective theory we find
\begin{eqnarray}
A_1 & = & -A_2\ =\ 
-\frac{\overline{\D}\D}{4 \pi^2} \log(\frac{\D}{\mu}) \ = \ A
\ \ ,\ \ A_3\ =\ 0
\ \ \ .
\end{eqnarray}
\noindent

The meson masses at leading order are found by expanding Eq.(\ref{EFT})
to second order in the meson fields.
The charged meson masses are
 \begin{eqnarray}
 \label{massesgen} 
 m_{\pi^+}^2 & = & {8A\over f_8^2}\  (m_u+m_d) m_s
\ \ ,\ \ 
 m_{K^+}^2 \  = \  {8A\over f_8^2}\  (m_u+m_s) m_d
\ \ ,\ \ 
m_{K^0}^2  =  {8A\over f_8^2}\  (m_d+m_s) m_u
\ \ \ ,
\end{eqnarray}
and the neutral meson mass matrix is
\begin{eqnarray}
\label{eq:neuts}
m_{33}^2 & = & {8A\over f_8^2}\  m_s(m_u+m_d)
\ \ ,\ \ 
m_{88}^2  =  {8A\over 3 f_8^2}
\left[ \  m_s(m_u+m_d)+4 m_u m_d\ \right]
\nonumber\\
m_{11}^2 & = & {16 A\over f_1^2}
\left[\ m_s (m_u+m_d) + m_u m_d\ \right]
\ \ ,\ \ 
m_{13}^2  =  -{8A\sqrt{2}\over{f_1 f_8}}\ (m_u-m_d) m_s
\nonumber\\
m_{18}^2 & = & {16A\over \sqrt{6}f_1 f_8}
\left[ \ m_s(m_u+m_d) - 2 m_u m_d \ \right]
\ \ ,\ \ 
m_{38}^2  = - {8A\over \sqrt{3} f_8^2}\ (m_u-m_d) m_s
\ \ \ .
\end{eqnarray}
\noindent
In the limit of zero mixing
$m_{11}=m_{\eta^\prime}$,
$m_{33}=m_{\pi^0}$ and 
$m_{88}=m_\eta$.
The eigenvectors of this matrix are close to the quark flavor eigenstates
for values of the quark masses consistent 
with standard chiral perturbation theory
\cite{GLclassic}.

It is interesting to consider mass relations between the mesons.
At this order in perturbation theory there is a relation between the 
meson masses without any assumption about the light quark 
mass hierarchy,
\begin{eqnarray}
m_{\eta^\prime}^2 + m_{\pi^0}^2 + m_{\eta}^2
& = & 
\left(\ {2\over 3}+{f_8^2\over f_1^2}\ \right)
\left[\   m_{\pi^+}^2 + m_{K^+}^2+ m_{K^0}^2
\ \right]
\ \ \ ,
\label{eq:massrel}
\end{eqnarray}
where $m_\eta$, $m_{\eta^\prime}$ and $m_{\pi^0}$ are the mass eigenvalues
found by diagonalizing the neutral meson mass matrix.
Further, if one assumes $m_{u,d}/m_s\ll 1$ and neglects such terms, 
\begin{eqnarray}
{m_{K^+}^2\over m_{\pi^+}^2} & = & {m_d\over m_d+m_u}
\ \ ,\ \ 
{m_{K^0}^2\over m_{\pi^+}^2}  =  {m_u\over m_d+m_u}
\ \ \ ,
\end{eqnarray}
and hence $m_{K^+}^2+m_{K^0}^2=m_{\pi^+}^2$.
It is also clear that there is an inverse mass hierarchy, e.g.
$m_{\pi^+}>  m_{K^+} > m_{K^0}$.

\begin{figure}[t]
\centerline{{\epsfxsize=3.0in \epsfbox{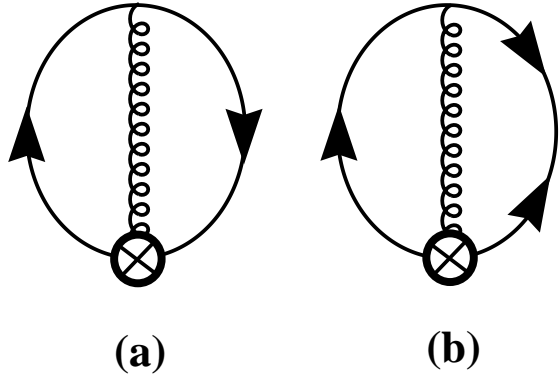}} }
\noindent
\caption{\it
Some of the two loop diagrams that contribute
to $A_3$.
The crossed circle denotes an insertion of a quark
mass operator involving the gluon field 
that arises at higher orders in the 
effective field theory expansion.
The curly lines denote gluons.
}
\label{fig:masscorr}
\vskip .2in
\end{figure}
In addition to corrections to the coefficients $A_1$ and $A_2$
arising at higher orders in $1/\mu$ and $\alpha_s$
there will be contributions to the $A_3$, such as those
shown in 
Fig.~(\ref{fig:masscorr}).
The operator in the effective theory appearing in 
Fig.~(\ref{fig:masscorr}) results from a two loop diagram
in QCD where two of the propagators are far off-shell, resulting
in an effectively local vertex for momenta much less than $\mu$.


\section{Two Flavors}

In the two flavor case (three colors)
the most favored condensate is
\begin{equation}
\langle \Psi^\a_{L a i} \Psi^\b_{L b j} \rangle = 
\langle \Psi^\a_{R a i} \Psi^\b_{R b j} \rangle = 
\D\epsilon^{\alpha\beta 3} \epsilon_{ab} \epsilon_{ij} 
\label{eq:twogap}
\end{equation}
\noindent
which breaks $SU(3)_c \otimes SU(2)_L\otimes SU(2)_R\otimes U(1)_A
\otimes U(1)_B\otimes U(1)_{em}$ down to $SU(2)_c \otimes SU(2)_L\otimes
SU(2)_R\otimes U(1)_{\rm \tilde{em}}\otimes U(1)_{\tilde{B}}$, where
$U(1)_{\rm \tilde{em}} $ and $U(1)_{\tilde{B}}$ are linear
combinations of electromagnetism, baryon number and the eighth gluon, 
$A^8_\mu$\cite{CFL_ARW,CFL_SW}.
At energies below the gap the dynamical degrees of freedom are 
three massless gluons, two ungapped quarks 
and the pseudo-Goldstone boson associated with spontaneous breaking of
$U(1)_A$.
In contrast to the three flavor case, there is no pseudo-Goldstone boson
associated with the breaking of baryon number.
The  Goldstone field describing the $\eta'$ meson related to the
breaking of $U(1)_A$ can be parametrized  as
\begin{eqnarray}
V & = & e^{i2 \eta'/f_{\eta'}}
\ \ ,\ \ 
V\rightarrow e^{-i 4 \beta} V
\ \ \ .
\end{eqnarray}
\noindent 
The effective lagrangian describing the dynamics of the 
$\eta^\prime$  at leading order 
in the $\partial/\D$ and $m/\D$ expansion is 
\begin{eqnarray}
{\cal L} &  = &  
{f_{\eta'}^2\over 8} \left(\ D_0 V D_0 V^*\ -\ |{\bf v}|^2 D_i V D_i V^*\ \right) 
\ +\ B\ \left[\  {\rm det}\left( \M  \right) V\ +\ 
{\rm h.c.} \ \right]
\ \ \ .
\end{eqnarray}
Using the diagonal basis for the gapped quark fields with $a,i=1,2$
\begin{equation}
\Psi^a_i =  {1\over\sqrt{2}} \sum_{A=1}^4
\left(\tau^A\right)^a_i \Psi^A
\end{equation}
\noindent
where $\tau^A$ are Pauli matrices for $A=1, 2,  3$ and
$\tau^4=I_2$, 
the gaps for the four gapped quarks are
$-\D_4=\D_A=\D $ for $A=1, 2, 3$, and the others remain ungapped.
Calculation of the decay constant and vacuum energy shift due to
the light quark masses is similar
to that performed in the three flavor case and gives
\begin{eqnarray}
f_{\eta'}^2 & = & \frac{\mu^2}{\pi^2}
\ \ ,\ \ 
|{\bf v}|^2\ =\ {1\over 3}
\ \ ,\ \ 
B\ =\ -{\overline{\D}\D\over {2\pi^2 }}\log(\frac{\D}{\mu})
\end{eqnarray}
\noindent
and therefore an $\eta'$ mass of
\begin{equation}
m^2_{\eta^\prime}  =  {{8{B_1}m_u m_d} \over{f^2_{\eta^\prime}}}=
-\frac{4\overline{\D}\D}{ \mu^2} \log(\frac{\D}{\mu})\  m_u m_d 
\ \ \ ,
\end{equation}
where we have neglected the contribution from the local counterterm
to the $\eta^\prime$ mass.

The ungapped quarks interact with themselves at leading order 
via the exchange of a massive gluon, inducing a four-quark operator
for scales below $g_s \mu$.  
The coefficient of this operator is independent of the strong coupling 
constant $\alpha_s$ (which arises from a cancellation between the couplings
and the gluon mass) and scales like $1/\mu^2$.
This interaction is repulsive since the two ungapped quarks 
are in a ${\bf 6}$ of color
and further the interaction vanishes in the high density limit. 
Thus we do not expect condensation of the ``green'' quarks in the 
high density limit.
It is interesting to note that 
the $\eta'$ does not couple directly to the ungapped quarks since
the $\eta'$ field is an excitation of a condensate involving the
``red'' and ``blue'' colors only.
Thus the axial coupling constant
describing this interaction is suppressed by powers of $\alpha_s$
and consequently the tensor force between the ungapped quarks
in the low energy theory
arising from the exchange of a single $\eta^\prime$
is suppressed by $\alpha_s^2$.

Another interesting aspect of the two-flavor case is the presence of an
unbroken pure $SU(2)_c$ gauge theory. 
The gluons associated with this gauge group do not interact with the 
ungapped quarks.
The confinement scale of this theory can
be estimated by assuming 
the only modification to the evolution of the strong coupling
arises from the particle content.
This provides an estimate of the scale at which the theory becomes strongly
coupled,
\begin{eqnarray}
\mu^{\rm conf.}_2 & = & 
\mu^{\rm conf.}_3\ \left({\Delta\over\Lambda_{\rm QCD}}\right)^{-{7\over 22}}
e^{6\pi \left({1\over 22}-{1\over 29}\right)}
\ \ \ ,
\end{eqnarray}
where $\mu^{\rm conf.}_{2,3}$ are these scales in the two and three
color theories respectively. 
For large $\Delta$, this scale is much lower than $\Lambda_{\rm QCD}$,
but for reasonable values of $\Delta$, 
$\mu^{\rm conf.}_2\sim \mu^{\rm conf.}_3$.
This suggests that pure Yang-Mills glueballs will
appear in the low energy theory with masses of order
$\sim\Lambda_{\rm QCD}$.

\section{Conclusion}

We have examined the high density limit of QCD 
where there are two and three flavors of ``light'' quarks 
below the scale relevant to the formation of a color superconducting state.
Using an effective field theory to describe quark modes near 
the Fermi surface we have determined the decay constants and masses
of the pseudo-Goldstone bosons that arise in each theory
at leading order in the $1/\mu$ expansion.
The masses of these pseudo-Goldstone modes vanish in the 
high density limit.
In order to determine the behavior of these systems at 
a moderate density the subleading corrections
(e.g. $1/\mu$, $\alpha_s$)
need to be determined, including the contributions
from instantons\cite{ManTyt}.

\vskip 1in

We would like to thank David Kaplan for encouragement.
This work is supported in part by the U.S. Dept. of Energy under
Grants No. DE-FG03-97ER4014 and DOE-ER-40561.

\vfill\eject

\end{document}